\documentclass[]{aastex631}

\usepackage{longtable}
\usepackage[normalem]{ulem}
\useunder{\uline}{\ul}{}
\usepackage{color, colortbl}
	
\usepackage{xcolor}

\shorttitle{Photobombing Earth 2.0}
\shortauthors{Saxena}
\graphicspath{{./}{figures/}}

\begin{document}

\title{Photobombing Earth 2.0: \\Diffraction Limit Related Contamination and Uncertainty in Habitable Planet Spectra}

\author{Prabal Saxena}
\email{prabal.saxena@nasa.gov}
\affiliation{CRESST II/University of Maryland, College Park, Maryland 20742, USA}
\affiliation{NASA Goddard Space Flight Center, Greenbelt, Maryland 20771, USA}



\begin{abstract}

Observing habitable exoplanets that may resemble Earth is a key priority in astronomy that is dependent on not only detecting such worlds, but also ascertaining that apparent signatures of habitability are not due to other sources.  Space telescopes designed to observe such worlds, such as that recommended by NASA's 2020 Astrophysics Decadal Survey, have a diffraction-limited resolution that effectively spreads light from a source in a region around the source point. In this letter, we show that the diffraction limit of a 6 meter space telescope results in a point spread function of an Earth-like planet that may contain additional unanticipated bodies for systems at distances relevant to proposed searches. These
unexpected additional objects, such as other planets and moons, can influence obtained spectra for a putative habitable planet by producing spurious features and adding additional uncertainty in the spectra.  A model of the Earth observed by a 6 meter space telescope as though it was an exoplanet shows that the light from the Earth would be blended with the Moon, Mercury, Venus and Mars in various combinations and at different times for numerous combinations of distance to the system and wavelength.  Given the importance of extricating the true spectra of a potentially habitable planet in order to search for biosignatures, we highlight the need to account for this effect during the development of relevant telescopes and suggest some potential means of accounting for this photobombing effect.  

\end{abstract}



\section{Introduction} \label{sec:intro}

Efforts to harness the power of direct imaging to explore habitable worlds is underlied by the belief that Earth is not only under our feet, but may also exist over our heads in other regions of space. Contextual information that requires both observational and theoretical reconnaissance of promising planets and their systems is critical to ensure a suspected habitable world deserves the resources required to characterize it.  Obtaining this contextual information must consider capabilities and limits of planned future telescopes at the forefront of exploring these high priority worlds.  In this study we discuss how the diffraction limit of these future telescopes may influence observation strategy and outcomes. A significant manner in which diffraction limits may do this is alluded to in the title, where 'photobombing' refers to an image that contains the unexpected appearance of an unintended object in the camera's field of view as the image was taken.  These objects may include other planets and moons, which can influence obtained spectra if they are unresolvable from the target planets' point-spread function (PSF). Understanding where/how such blends may occur is important in development of future telescopes aimed at detecting habitable exoplanets, such as the recently recommended infrared/optical/ultraviolet telescope in the 2020 Astrophysical Decadal Survey \citep{2021pdaa.book.....N}.  Recent frameworks for assessing potential biosignatures \citep{2018AsBio..18..709C, 2021Natur.598..575G} have stressed the need to first extricate a true signature potentially indicative of life and then vet potential complicating scenarios that may confound accurate interpretation of a signature, and we will show that photobombing by additional planets or moons may complicate both those requirements.

Exploration of this effect deserves a thorough treatment as there will likely be variation due to range of properties relevant to specific observations, including system inclination, planetary orbital phase, intrinsic spatial variations in each relevant body and other system and planetary characteristics.  However, this letter is an initial exploratory study focusing on four topics that guide further study and constitute the paper's sections.  First, we examine the angular size of the diffraction limit of several telescope diameter/observational wavelength combinations as a function of distance to targeted systems.  This includes comparison to key distances within potential planetary systems that may exist around stars of different stellar types - the habitable zone width (HZ) and size of the Hill radius for a mid-HZ Earth-twin.  Second, we use our solar system as a model for the type of photobombing effect that may be observed for planetary systems with multiple small, rocky planets in/near the habitable zone. We examine the likelihood that additional planets appear in the Earth's PSF as a function of time, if our solar system was observed using a 6 meter telescope.  Third, we model consequences such contamination of the Earth's PSF would have on snapshot visible and near-infrared spectra obtained for the Earth using the telescope from 10 parsecs away.  Finally, we discuss other potential photobombing scenarios relevant to future observations and potential means of extricating the true Earth-twin spectrum from potential contamination.

\section{Diffraction Limit Comparisons for Future Telescopes} \label{sec:Compare}

The \textcolor{black}{theoretical} minimum diffraction-limited angular resolution a telescope can resolve is related to the wavelength of observation and the diameter of the entrance aperture of the primary objective, typically the telescope mirror. For space telescopes that aim to observe potentially rocky planets in a system's habitable zone in visible or near-infrared light, this contributes to/helps set limits on both the minimum Inner Working Angle (IWA) (the $\lambda/d$ distance at which off-axis throughput of a coronagraphic system $>$ 50\%, where $\lambda$ is the wavelength of interest and d is the diameter of the entrance aperture of the primary objective - \textcolor{black}{note that this is mask dependent}) and angular size of the point spread function of an object of interest that exceeds the contrast floor for the telescope \citep{2010exop.book..111T}. This  heavily influences yields for a given telescope that is attempting to observe specific classes of planets \citep{2019JATIS...5b4009S}.

\begin{figure}[h!]
    \centering
    \includegraphics[angle=0]{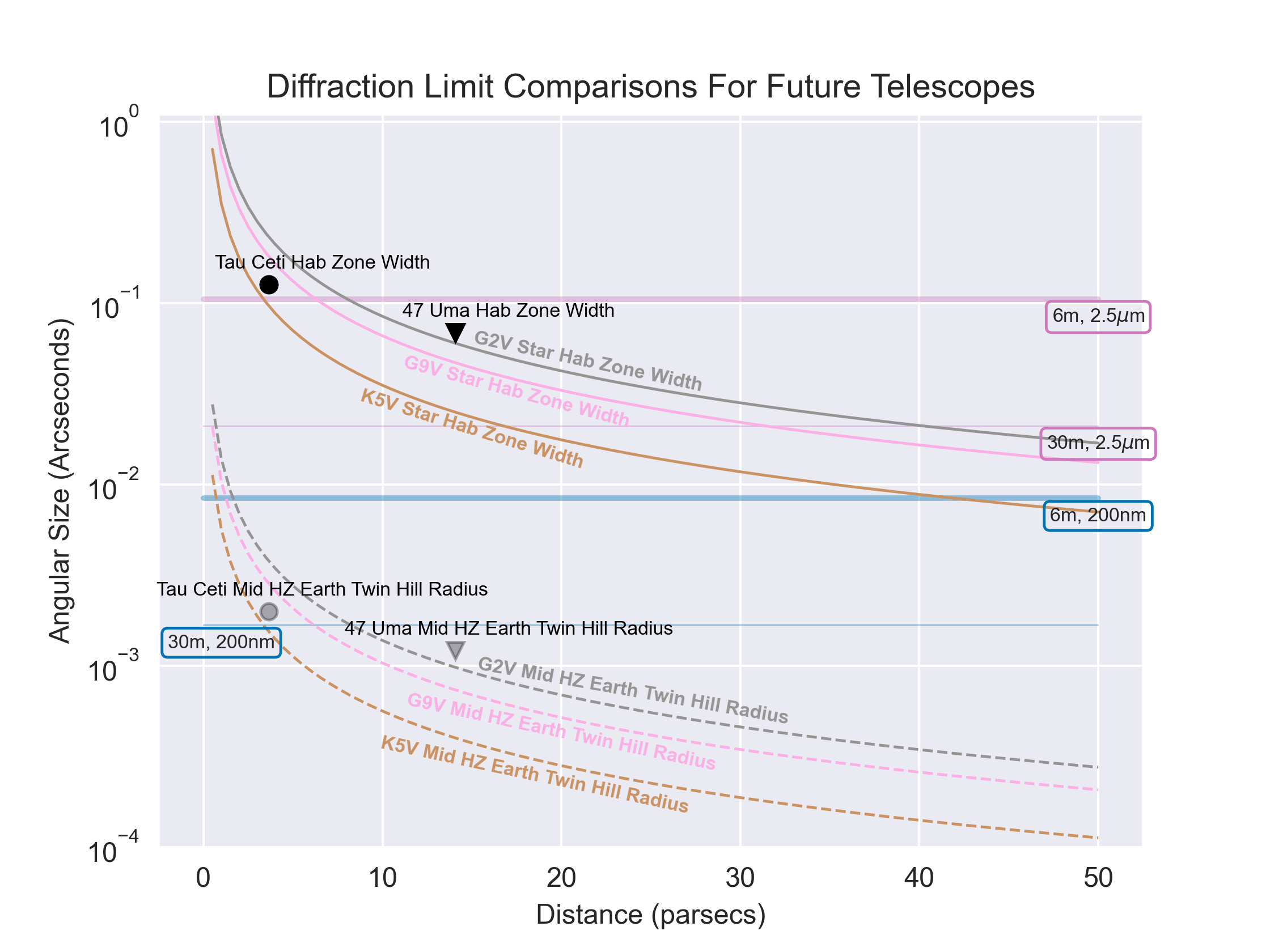}
    \caption{ Diffraction limit comparisons for future telescopes of 6 and 30 meter mirror size at 200 nm and 2.5 $\mu m$ versus Habitable Zone widths and Mid-HZ Earth-twin Hill Radii for G2V, G9V, K5V stars as a function of distance.   }
        \label{figure:DiffractionLimit}
\end{figure}

The diffraction limit of a telescope also influences characterization of a target because minimum PSF size corresponds to a physical distance below which additional sources (such as additional planets/a moon) cannot be resolved. Previous studies have demonstrated an unresolved moon may contribute to an obtained spectra for a planet in ways that complicate interpretation and produce false positive molecular detections \citep{2011ApJ...741...51R, 2014PNAS..111.6871R}.  We examine how the diffraction limit for a hypothetical Decadal-like 6m space telescope may influence PSF size for targets of interest for a range of potential planetary systems with different host stars at different distances.  In figure \ref{figure:DiffractionLimit}, we compare diffraction limits to two distances for which we are concerned with 'photobombing' - the Hill radius of an Earth-twin planet relevant to moons, and the HZ width for the system, relevant to other potentially unresolvable planets within the PSF. \textcolor{black}{Note that this theoretical diffraction limited PSF size is optimistic since coronagraph masks can broaden PSF size by, for example, effectively decreasing the pupil diameter.(private communication, Neil Zimmerman)}

We calculated the diffraction limits (blue/red horizontal lines in figure \ref{figure:DiffractionLimit}) using the formula for the first zero-intensity angle for a clear circular aperture ($\approx 1.22 \lambda/d$), which gives an estimate of the distance over which an additional object may not be distinguishable within a PSF.  However, intensity drops off from PSF's central peak with a full-width at half-maximum at $\approx 80\%$ of that value.  This may have implications on extricating potential photobombing signals (see section \ref{sec:discuss}). For reference wavelengths, we use minimum/maximum wavelengths from two direct imaging mission studies commissioned for the Decadal: HabEx \citep{2020arXiv200106683G} and LUVOIR \citep{2019arXiv191206219T}, which corresponds to 200 nm and 2.5 $\mu m$ for the Extreme Coronagraph for Living Planetary Systems instrument on LUVOIR.  We modeled the diffraction limit and other distance calculations for systems out to 50 parsecs, based on the cutoff used to maximize Exo-Earth yield in a previous study \citep{2014ApJ...795..122S}. 

We compared the diffraction limit for a Decadal-like telescope to the HZ width and to the Hill Sphere for a mid-HZ Earth-twin in the system. 
Figure \ref{figure:DiffractionLimit} assumes the most optimal observing case \textcolor{black}{for avoiding photobombing}, a planet-star system observed face on, and zero planetary eccentricity or inclination. The middle, boundary and total width values of the HZ for different stars was taken from figure 7 of \citet{2013ApJ...765..131K}. HZ and Hill Sphere widths were plotted for a G2V, G9V and K5V star.  They were also calculated for smaller stars, but the distance to the middle of the M1V HZ is interior to the IWA plus diffraction limit in the blue optical (450 nm) for distances $\approx >5 pc$ so they were not included. We calculated the Hill Sphere for a simplified case where the planetary orbital eccentricity is 0 \citep{1992Icar...96...43H} using the equation $R_{H} = a * (M_{planet}/3M_{star})^{1/3}$, where a is semi-major axis and $M_{planet}$ and $M_{star}$ are mass of the planet and star.  The Hill radius calculated using this equation is significantly larger than estimates of true regions of long term stability for prograde moons around a planet ($\approx0.36R_{H}$ \citep{2002ApJ...575.1087B}).  Finally we annotated figure \ref{figure:DiffractionLimit} with HZ width and Hill Radius values for two close planet hosting systems with stars similar to the Sun's spectral class. We also included diffraction limits for a 30m telescope for reference.

There are several implications of the comparisons in figure \ref{figure:DiffractionLimit}:

\begin{enumerate}
    \item The diffraction limit for a 6m class telescope is larger than HZ width for Sun-like and smaller stars in the infrared for many of the distances examined.  While the diffraction limit for 2.5 $\mu m$ is clearly larger than the HZ for all three stars at distances $\approx >10 pc$, the diffraction limit for 1.0 $\mu m$ is $\approx$ equal to HZ width for the G9V and G2V system at $\approx$ 25 and 35 pc, respectively. Note the diffraction limit is the one-sided distance from central peak to first minima in the Airy disk, so the more appropriate comparison for PSF width versus HZ width may be 2*diffraction limit, which suggests even closer distances would have commensurate sizes for these two values. For a 30m telescope, the diffraction limit does not equal HZ width until about 50 pc away for 2.5 $\mu m$, thus next generation large ground telescopes with adaptive optics may potentially aid in constraining planet position. 
     \item For a mid-HZ target in a case where the diffraction limit size $\approx$ HZ width, the PSF would contain additional habitable zone objects that fall within the PSF in the direction of orbital motion as well. For cases where the diffraction limit is larger than HZ width or the target is closer to a HZ edge with a significant PSF size, potential photobombing planets may include those outside the HZ and unless there are ideal observational conditions, there may be significant uncertainty in the targets' estimated orbital separation.  The climate of planets is sensitive to orbital separation as it determines incoming stellar flux received - supplementary orbit constraining measurements using radial velocity and astrometry may be critical.
     \item For a decadal-like telescope, at basically all wavelengths at relevant distances examined, the Hill sphere for a mid-HZ Earth-twin is significantly smaller than the diffraction limit.  This reinforces previous studies' findings that a bound moon would be difficult to image from a planet, particularly since bound moons are likely to exist at a fraction of the hill radius distance. Only for the closest distances at the bluest wavelengths does the 30m telescope come close to being able to probe the hill sphere.
\end{enumerate}

\begin{figure}[b!]
    \centering
    \includegraphics[angle=0, scale = 0.54]{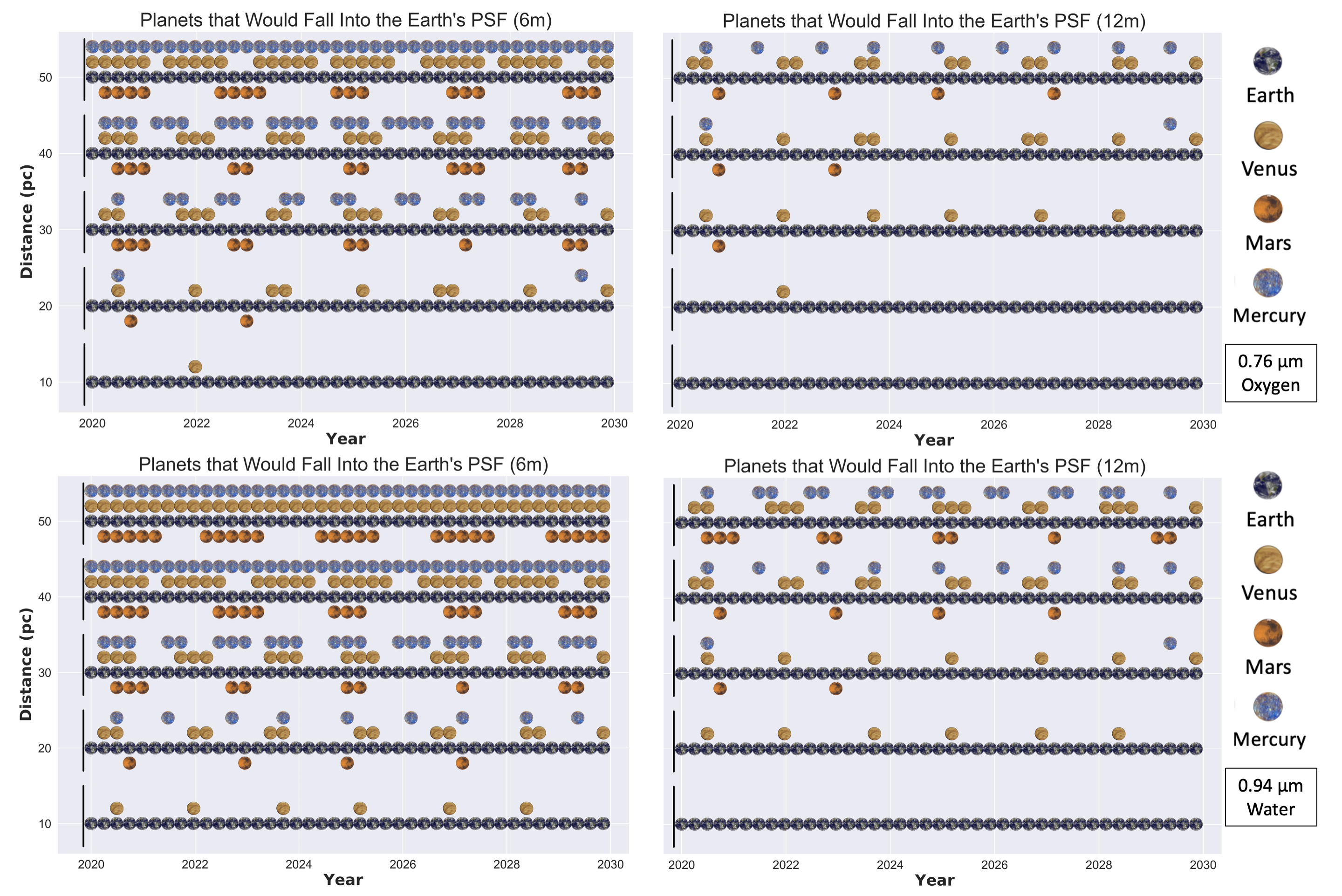}
    \caption{  Time series figures displaying which planets would fall into the Earth's PSF from 2020-2030 if observed by a 6 and 12 meter telescope in the 0.76 $\mu m$/0.94 $\mu m$ oxygen/water bands from different distances. \textcolor{black}{Total yearly photobombing duration for each planet is given in table \ref{tab:Photodays}.}   }
        \label{figure:Earth1pt0}
\end{figure}

\section{Photobombing in Our Solar System} \label{sec:SSphotobomb}

Our calculations shown in figure \ref{figure:DiffractionLimit} showed that the diffraction limit of a 6m telescope may result in large PSF sizes compared to HZ widths.  We calculated how our solar system and the Earth's PSF would look if it were observed by a similar telescope at different distances in the range used to generate a sufficient Exo-Earth yield. We examined what planets would fall into the Earth's PSF based upon their time variable distances to the Earth from 2020-2030. We counted any planet closer to the Earth than the diffraction limit distance as a 'photobombing' planet. We used the Skyfield program \citep{2019ascl.soft07024R} to calculate daily planet positions and distance to Earth for Mercury, Venus, and Mars.  The Moon always falls into the Earth's PSF for the distances we examined.  

Figure \ref{figure:Earth1pt0} shows what planets fall into the Earth's diffraction-limited PSF using a 6m telescope and also 12m telescope from 10, 20, 30, 40 and 50 parsecs away, \textcolor{black}{while table \ref{tab:Photodays} shows how many days each planet would photobomb Earth in each year}.  The top/bottom panels \textcolor{black}{of figure \ref{figure:Earth1pt0} and sections of table \ref{tab:Photodays}} correspond to the 0.76 $\mu m$/0.94 $\mu m$ oxygen and water absorption features, which are of high interest for determining habitability and similarity to Earth. Which planets photobomb a particular PSF is wavelength dependent.  The Earth is always in the PSF and is the constant line across plots at different distances, with Mercury, Venus and Mars displayed next to the Earth if they are close enough be blended into the Earth's PSF at that time. The figures based on assumptions that the system is being viewed face on \textcolor{black}{(the best case scenario in terms of limiting the duration of photobombing of Earth)}, and that anything that falls into the diffraction limit distance is considered a photobomb, regardless of whether it may be visible relative to the drop off in flux from the central peak.  Each vertical set of planet points along the x-axis \textcolor{black}{in figure \ref{figure:Earth1pt0}} corresponds to a date 90 days from the previous one, during which a planet may move in/outside the PSF - this photobombing frequency corresponds strongly to total time a planet falls into the Earth's PSF, \textcolor{black}{as demonstrated by total photobombing times given in table \ref{tab:Photodays}}. 

Key points from this analysis include:

\begin{enumerate}
     \item Photobombing of Earth by neighboring planets is a common occurrence past 10 parsecs for a 6m telescope. There are $<30$ stars within 10 parsecs of a somewhat similar stellar type as the Sun.  If Exo-Earths exist in systems similar to our own, it will be common to have additional planets contained in an Exo-Earth's PSF \textcolor{black}{at numerous wavelengths of interest} at distances exceeding 15 parsecs. This changes for a 12m telescope, which has the angular resolution to distinguish many photobombing planets.  
     \item Photobombing of Earth by neighboring planets observed from $>20$ parsecs and farther is persistent over time. In the water band at 20 parsecs, an additional planet photobombs Earth's PSF $\approx 50\%$ of the time \textcolor{black}{(See table \ref{tab:Photodays} for individual planet durations)}. This is lower at shorter wavelengths, but more common for longer wavelengths relevant to critical NIR signatures. At larger distances of observation, photobombing should be considered a potential regular occurrence over time.
     \item  Time varying multi-wavelength observations may be able to extricate photobombing signatures, but are complicated by the likelihood multiple and different planets may be in the Earth's PSF simultaneously \textcolor{black}{and that photobombing duration from one planet can vary from year to year}.  Additionally, the Moon will continuously photobomb the Earth, and each of the bodies in the PSF may exhibit time variable spectra due to intrinsic spatial variation of bodies' spectra. This is further complicated if realistic system and planetary properties such as phase and inclination varying albedo are included. 
\end{enumerate}


\section{Consequences for Obtained Spectra} \label{sec:spectra}

The possibility of Exo-Earth PSF contamination by other bodies necessitates understanding of the potential effect on obtained spectra.  We extend our Earth as an exoplanet analogue to probe this effect by simulating photobombing spectra scenarios using the Planetary Spectrum Generator (PSG) \citep{2018JQSRT.217...86V}.  We simulated spectra of the inner solar system planets and Moon as observed by a 6m space telescope using a telescope/instrument configuration similar to the LUVOIR ECLIPS mode.  We adapted models and observing parameters from \citet{2021AJ....161..150C} to simulate spectra of the system as viewed edge on (system inclination = $90^{\circ}$) from 10 parsecs away with the planets at quadrature. Sampling techniques and radiative transfer details are given in \citet{2021AJ....162...30S} and observation geometry, planet, and instrument properties are in config files (see supplemental).    

Simulated planet spectra assumed a 1000 second maximum exposure time based on current and planned missions' attempts to minimize impact of cosmic ray interference \citep{10.1117/12.2563480, 10.2307/26874443}. Total observation time was set to be equal for all planets to probe the photobombing effects and set to 1000 hours, the approximate time required to distinguish key features in the Earth's visible spectra to $\approx5\sigma$.  This is also the approximate time of a single continuous starshade observability window assumed for the HabEx mission \citep{2020arXiv200106683G} and roughly half the time of the x-axis spacing in figure \ref{figure:Earth1pt0}. 

\begin{figure}[t!]
    \centering
    \includegraphics[angle=0, scale = 0.80]{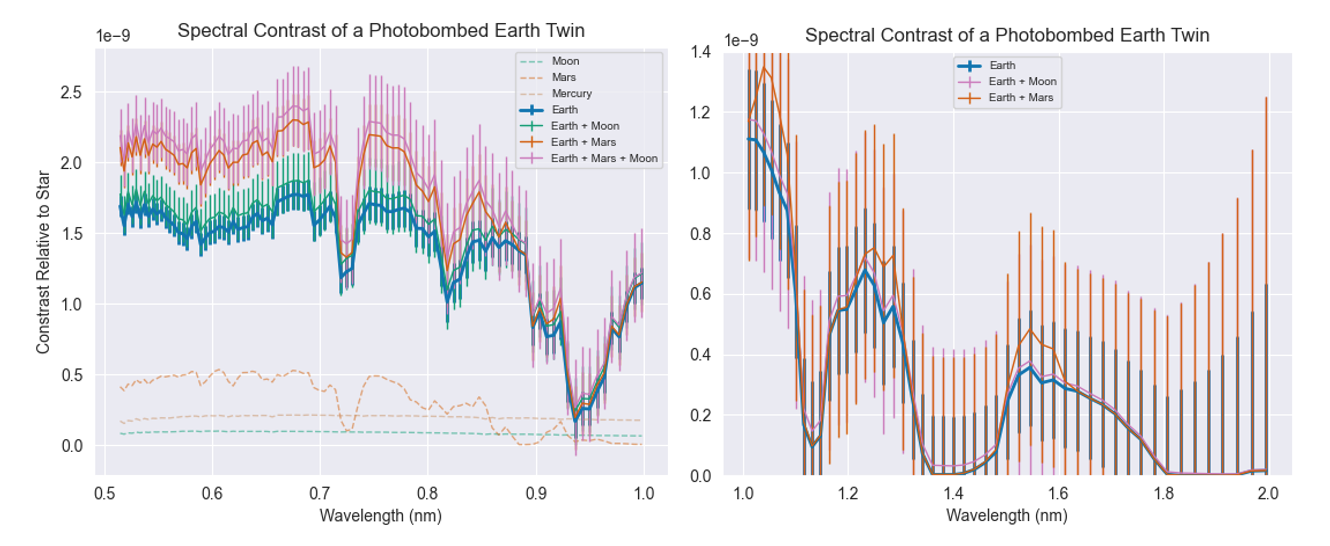}
    \caption{ Figures of spectral contrast in the visible (left) and near-infrared (right) of a photobombed Earth-twin for different scenarios. Noise assumptions given in config files. }
        \label{figure:Spectra}
\end{figure}

Results are in figure \ref{figure:Spectra} for the visible/near-infrared (left/right) wavelength ranges of ECLIPS.  We plot individual planet spectra and spectra (with noise) of Earth alone and the Earth with three other photobombing scenarios in the visible, and Earth spectra (with noise) and two other photobombing scenarios for the near-infrared. The most notable feature of photobombing scenarios in the visible (Earth + Moon, Earth + Mars and Earth + Mars + Moon) is that in each case the additional source in the PSF results in additional flux and/or noise that complicates interpretation of spectra with respect to extricating molecular signatures.  In the Earth + Moon scenario, overall continuum flux level is barely changed relative to the Earth only case, but oxygen feature depth is reduced and there is an increase in noise that reduces the significance of multiple features. In other visible spectra photobombing scenarios, the continuum flux and noise goes up significantly.  While this reduces significance of some molecular features, there is also an additional spurious increase in the oxygen signature absorption depth, which may result in an erroneously interpreted greater oxygen abundance. In the infrared spectra, photobombing cases have some minor effects on continuum and molecular signature flux (with the Earth + Mars case having larger effects), but the greatest effect is significant increase in noise.  This increase reduces the significance of molecular signatures (particularly in the Earth + Mars case) such that robust detection may not be possible in such observing scenarios.  Note these simplified scenarios provide a snapshot of effects on spectra, without phase and inclination effects taken into account, and no consideration of other factors that may cause time varying spectra.  They are also dependent on observational assumptions listed above.

\section{Mitigation Strategies and Discussion} \label{sec:discuss}

\begin{figure}[t!]
    \centering
    \includegraphics[angle=0, scale = 0.90]{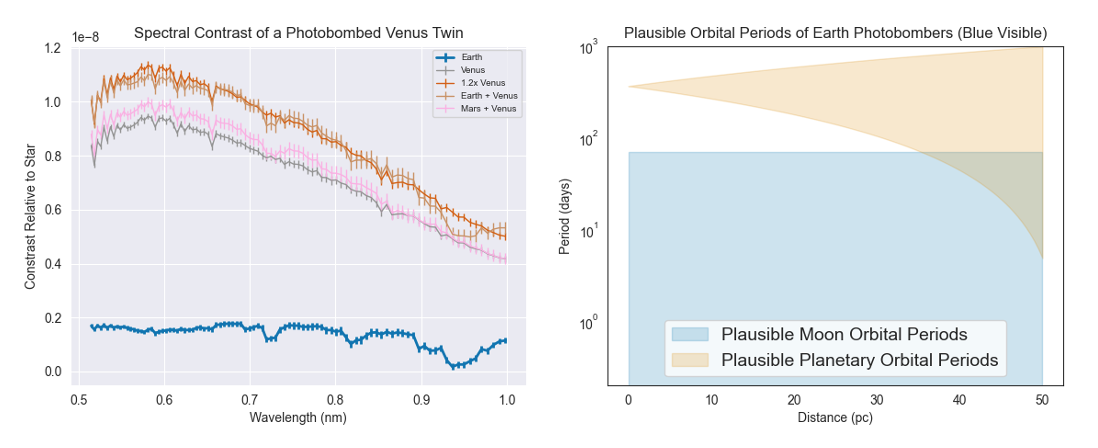}
    \caption{Spectral contrast of a photobombed Venus Twin for different scenarios (left) and plausible orbital periods for potential photobombers of a Earth-twin in the blue visible from 10 parsecs away (right).}  
        \label{figure:VenusExclusion}
\end{figure}

The unexpected presence of planets and moons in the PSF of a planet targeted in a search for an Exo-Earth complicates accurate assessment of planetary properties for future efforts.  For a 6m telescope such as the Astrophysics Decadal Survey-like flagship telescope, the diffraction-limited PSF of a target may be greater than habitable zone width for systems of interest and our exploration of how our solar system might appear to such a telescope indicates photobombing effects may be a complicated time variable phenomenon. Contamination of target planets' obtained spectra by other bodies can induce significant changes in flux levels and noise that produce muted or erroneous spectral signatures.  This is more complicated when time varying variations due to variations in both target planet and photobombing bodies are considered (e.g. spatial variations, weather, megastructures, etc).  Such photobombing effects may also hide existence of interesting, potentially habitable planets.  The left panel of figure \ref{figure:VenusExclusion} shows simulated spectra for a Venus Twin of different sizes that is photobombed by additional planets. Note that with possible exception of the water band, it is difficult to tell apart a $1.2R_{Venus}$ Venus-like planet and a blended Venus and Earth spectra.  Both scenarios indicate the ability to extricate potential additional bodies from a target planets' PSF should be incorporated into planning for targeting of Exo-Earths, since the presence of multiple small planets around such worlds, while uncertain, may be a possibility \citep{2019MNRAS.483.4479Z, 2020AJ....159..248K}.  \textcolor{black}{Mitigation strategies may be considered through both observatory design considerations and other methods.}

\textcolor{black}{\textbf{Observatory Design Based Mitigation Strategies}: One way of obviating photobombing concerns is increased mirror size - figure \ref{figure:Earth1pt0} and table \ref{tab:Photodays} show the improvement a 12m telescope offers.  However, increasing mirror size comes with concerns regarding practical trades/programmatic concerns, which future work should explore.} Image plane Nyquist sampling for future detectors on such telescopes may need to be modified to ensure such issues are tractable for certain standard systems at pre-determined distances to ensure a sufficient sample of systems where photobombers can be identified.  \textcolor{black}{\textbf{Other Mitigation Strategies}: Detector image plane sampling would be important for mitigation using} relative fluxes of photobombers and their location versus PSF peak and dropoff in PSF intensity - this was ignored for this initial exploratory work, but should be studied. Additional mitigation techniques may apply to certain observing scenarios, such as occurrence of serendipitious mutual occultations \citep{2007A&A...464.1133C} or use of spectroastrometry \citep{2015ApJ...812....5A} (using multiple telescopes with different mirror sizes/spectral ranges). Multi-epoch observations supported by theoretical exploration of likely signatures may help address such scenarios.  The right panel of figure \ref{figure:VenusExclusion} shows plausible moon orbital periods around an Earth-twin by calculating Roche Limit and 1/3 * Hill Sphere bounded Keplerian periods and plausible orbital periods of a photobombing planet contained within a distance-to-system dependent blue-visible (450nm) diffraction limit to produce zones where time variable signatures may be due to a photobomber. System specific modeling can incorporate such limits to determine optimal sampling of multi-epoch observations. \textcolor{black}{Many strategies/techniques might be tested using the Nancy Grace Roman Telescope's Coronagraph Instrument \citep{2019arXiv190104050B} in a different phase space, to help reduce photobombing risk for future telescopes or suggest mirror size is the critical means of addressing the issue.} Finally, supplementary and synergistic high resolution spectroscopy \citep{2015A&A...576A..59S} or interferometry \citep{2019A&A...623L..11G} observations may help address photobombing observations before additional potentially diagnostic missions arrive \citep{2021arXiv210107500Q, 2022arXiv220404866T}.

\begin{acknowledgments}

Acknowledgements: We thank Thomas Fauchez, Avi Mandell, Shawn Domagal-Goldman and Daya Saxena for their helpful conversations that improved the quality of this manuscript. \textcolor{black}{We'd also like to thank the reviewer, whose suggestions improved the quality of this paper.} This work was supported by NASA under award number 80GSFC21M0002 and was also funded in part by the GSFC Sellers Exoplanet Environments Collaboration (SEEC).  

\end{acknowledgments}
\newpage
\begin{longtable}{|ccccccccccc|}
\caption{Number of days each planet photobombs Earth each year, where * represents the entire year.}
\label{tab:Photodays}\\
\hline
\multicolumn{1}{|c|}{\textbf{\begin{tabular}[c]{@{}c@{}}Planet, Mirror\\ Size, Distance\end{tabular}}} & \multicolumn{1}{c|}{\textbf{2020}} & \multicolumn{1}{c|}{\textbf{2021}} & \multicolumn{1}{c|}{\textbf{2022}} & \multicolumn{1}{c|}{\textbf{2023}} & \multicolumn{1}{c|}{\textbf{2024}} & \multicolumn{1}{c|}{\textbf{2025}} & \multicolumn{1}{c|}{\textbf{2026}} & \multicolumn{1}{c|}{\textbf{2027}} & \multicolumn{1}{c|}{\textbf{2028}} & \textbf{2029} \\ \hline
\endfirsthead
\endhead
\multicolumn{11}{|c|}{{\ul \textbf{Oxygen (0.76 microns)}}} \\ \hline
\multicolumn{1}{|c|}{\textbf{Mercury/6m/10pc}} & \multicolumn{1}{c|}{0} & \multicolumn{1}{c|}{0} & \multicolumn{1}{c|}{0} & \multicolumn{1}{c|}{0} & \multicolumn{1}{c|}{0} & \multicolumn{1}{c|}{0} & \multicolumn{1}{c|}{0} & \multicolumn{1}{c|}{0} & \multicolumn{1}{c|}{0} & 0 \\ \hline
\multicolumn{1}{|c|}{\textbf{Mercury/6m/20pc}} & \multicolumn{1}{c|}{30} & \multicolumn{1}{c|}{26} & \multicolumn{1}{c|}{26} & \multicolumn{1}{c|}{29} & \multicolumn{1}{c|}{31} & \multicolumn{1}{c|}{33} & \multicolumn{1}{c|}{32} & \multicolumn{1}{c|}{25} & \multicolumn{1}{c|}{26} & 24 \\ \hline
\multicolumn{1}{|c|}{\textbf{Mercury/6m/30pc}} & \multicolumn{1}{c|}{142} & \multicolumn{1}{c|}{141} & \multicolumn{1}{c|}{150} & \multicolumn{1}{c|}{158} & \multicolumn{1}{c|}{152} & \multicolumn{1}{c|}{143} & \multicolumn{1}{c|}{141} & \multicolumn{1}{c|}{141} & \multicolumn{1}{c|}{141} & 155 \\ \hline
\multicolumn{1}{|c|}{\textbf{Mercury/6m/40pc}} & \multicolumn{1}{c|}{251} & \multicolumn{1}{c|}{259} & \multicolumn{1}{c|}{269} & \multicolumn{1}{c|}{269} & \multicolumn{1}{c|}{269} & \multicolumn{1}{c|}{260} & \multicolumn{1}{c|}{250} & \multicolumn{1}{c|}{251} & \multicolumn{1}{c|}{263} & 268 \\ \hline
\multicolumn{1}{|c|}{\textbf{Mercury/6m/50pc}} & \multicolumn{1}{c|}{*} & \multicolumn{1}{c|}{*} & \multicolumn{1}{c|}{*} & \multicolumn{1}{c|}{*} & \multicolumn{1}{c|}{*} & \multicolumn{1}{c|}{*} & \multicolumn{1}{c|}{*} & \multicolumn{1}{c|}{*} & \multicolumn{1}{c|}{*} & * \\ \hline
\multicolumn{1}{|c|}{\textbf{Venus/6m/10pc}} & \multicolumn{1}{c|}{28} & \multicolumn{1}{c|}{12} & \multicolumn{1}{c|}{29} & \multicolumn{1}{c|}{29} & \multicolumn{1}{c|}{0} & \multicolumn{1}{c|}{33} & \multicolumn{1}{c|}{37} & \multicolumn{1}{c|}{0} & \multicolumn{1}{c|}{28} & 12 \\ \hline
\multicolumn{1}{|c|}{\textbf{Venus/6m/20pc}} & \multicolumn{1}{c|}{123} & \multicolumn{1}{c|}{59} & \multicolumn{1}{c|}{75} & \multicolumn{1}{c|}{125} & \multicolumn{1}{c|}{0} & \multicolumn{1}{c|}{128} & \multicolumn{1}{c|}{130} & \multicolumn{1}{c|}{0} & \multicolumn{1}{c|}{123} & 59 \\ \hline
\multicolumn{1}{|c|}{\textbf{Venus/6m/30pc}} & \multicolumn{1}{c|}{207} & \multicolumn{1}{c|}{101} & \multicolumn{1}{c|}{116} & \multicolumn{1}{c|}{208} & \multicolumn{1}{c|}{27} & \multicolumn{1}{c|}{186} & \multicolumn{1}{c|}{171} & \multicolumn{1}{c|}{43} & \multicolumn{1}{c|}{207} & 101 \\ \hline
\multicolumn{1}{|c|}{\textbf{Venus/6m/40pc}} & \multicolumn{1}{c|}{305} & \multicolumn{1}{c|}{146} & \multicolumn{1}{c|}{161} & \multicolumn{1}{c|}{288} & \multicolumn{1}{c|}{92} & \multicolumn{1}{c|}{232} & \multicolumn{1}{c|}{215} & \multicolumn{1}{c|}{91} & \multicolumn{1}{c|}{306} & 146 \\ \hline
\multicolumn{1}{|c|}{\textbf{Venus/6m/50pc}} & \multicolumn{1}{c|}{366} & \multicolumn{1}{c|}{218} & \multicolumn{1}{c|}{224} & \multicolumn{1}{c|}{360} & \multicolumn{1}{c|}{222} & \multicolumn{1}{c|}{299} & \multicolumn{1}{c|}{281} & \multicolumn{1}{c|}{223} & \multicolumn{1}{c|}{366} & 218 \\ \hline
\multicolumn{1}{|c|}{\textbf{Mars/6m/10pc}} & \multicolumn{1}{c|}{0} & \multicolumn{1}{c|}{0} & \multicolumn{1}{c|}{0} & \multicolumn{1}{c|}{0} & \multicolumn{1}{c|}{0} & \multicolumn{1}{c|}{0} & \multicolumn{1}{c|}{0} & \multicolumn{1}{c|}{0} & \multicolumn{1}{c|}{0} & 0 \\ \hline
\multicolumn{1}{|c|}{\textbf{Mars/6m/20pc}} & \multicolumn{1}{c|}{121} & \multicolumn{1}{c|}{0} & \multicolumn{1}{c|}{64} & \multicolumn{1}{c|}{1} & \multicolumn{1}{c|}{0} & \multicolumn{1}{c|}{0} & \multicolumn{1}{c|}{0} & \multicolumn{1}{c|}{0} & \multicolumn{1}{c|}{0} & 0 \\ \hline
\multicolumn{1}{|c|}{\textbf{Mars/6m/30pc}} & \multicolumn{1}{c|}{205} & \multicolumn{1}{c|}{8} & \multicolumn{1}{c|}{121} & \multicolumn{1}{c|}{41} & \multicolumn{1}{c|}{51} & \multicolumn{1}{c|}{73} & \multicolumn{1}{c|}{4} & \multicolumn{1}{c|}{109} & \multicolumn{1}{c|}{0} & 123 \\ \hline
\multicolumn{1}{|c|}{\textbf{Mars/6m/40pc}} & \multicolumn{1}{c|}{251} & \multicolumn{1}{c|}{41} & \multicolumn{1}{c|}{179} & \multicolumn{1}{c|}{74} & \multicolumn{1}{c|}{96} & \multicolumn{1}{c|}{107} & \multicolumn{1}{c|}{41} & \multicolumn{1}{c|}{147} & \multicolumn{1}{c|}{2} & 201 \\ \hline
\multicolumn{1}{|c|}{\textbf{Mars/6m/50pc}} & \multicolumn{1}{c|}{291} & \multicolumn{1}{c|}{74} & \multicolumn{1}{c|}{238} & \multicolumn{1}{c|}{106} & \multicolumn{1}{c|}{151} & \multicolumn{1}{c|}{142} & \multicolumn{1}{c|}{80} & \multicolumn{1}{c|}{187} & \multicolumn{1}{c|}{35} & 255 \\ \hline
\multicolumn{1}{|c|}{\textbf{Mercury/12m/10pc}} & \multicolumn{1}{c|}{0} & \multicolumn{1}{c|}{0} & \multicolumn{1}{c|}{0} & \multicolumn{1}{c|}{0} & \multicolumn{1}{c|}{0} & \multicolumn{1}{c|}{0} & \multicolumn{1}{c|}{0} & \multicolumn{1}{c|}{0} & \multicolumn{1}{c|}{0} & 0 \\ \hline
\multicolumn{1}{|c|}{\textbf{Mercury/12m/20pc}} & \multicolumn{1}{c|}{0} & \multicolumn{1}{c|}{0} & \multicolumn{1}{c|}{0} & \multicolumn{1}{c|}{0} & \multicolumn{1}{c|}{0} & \multicolumn{1}{c|}{0} & \multicolumn{1}{c|}{0} & \multicolumn{1}{c|}{0} & \multicolumn{1}{c|}{0} & 0 \\ \hline
\multicolumn{1}{|c|}{\textbf{Mercury/12m/30pc}} & \multicolumn{1}{c|}{0} & \multicolumn{1}{c|}{0} & \multicolumn{1}{c|}{0} & \multicolumn{1}{c|}{0} & \multicolumn{1}{c|}{0} & \multicolumn{1}{c|}{0} & \multicolumn{1}{c|}{0} & \multicolumn{1}{c|}{0} & \multicolumn{1}{c|}{0} & 0 \\ \hline
\multicolumn{1}{|c|}{\textbf{Mercury/12m/40pc}} & \multicolumn{1}{c|}{30} & \multicolumn{1}{c|}{26} & \multicolumn{1}{c|}{26} & \multicolumn{1}{c|}{29} & \multicolumn{1}{c|}{31} & \multicolumn{1}{c|}{33} & \multicolumn{1}{c|}{32} & \multicolumn{1}{c|}{25} & \multicolumn{1}{c|}{26} & 24 \\ \hline
\multicolumn{1}{|c|}{\textbf{Mercury/12m/50pc}} & \multicolumn{1}{c|}{95} & \multicolumn{1}{c|}{94} & \multicolumn{1}{c|}{97} & \multicolumn{1}{c|}{111} & \multicolumn{1}{c|}{97} & \multicolumn{1}{c|}{97} & \multicolumn{1}{c|}{95} & \multicolumn{1}{c|}{95} & \multicolumn{1}{c|}{93} & 102 \\ \hline
\multicolumn{1}{|c|}{\textbf{Venus/12m/10pc}} & \multicolumn{1}{c|}{0} & \multicolumn{1}{c|}{0} & \multicolumn{1}{c|}{0} & \multicolumn{1}{c|}{0} & \multicolumn{1}{c|}{0} & \multicolumn{1}{c|}{0} & \multicolumn{1}{c|}{0} & \multicolumn{1}{c|}{0} & \multicolumn{1}{c|}{0} & 0 \\ \hline
\multicolumn{1}{|c|}{\textbf{Venus/12m/20pc}} & \multicolumn{1}{c|}{28} & \multicolumn{1}{c|}{12} & \multicolumn{1}{c|}{29} & \multicolumn{1}{c|}{29} & \multicolumn{1}{c|}{0} & \multicolumn{1}{c|}{33} & \multicolumn{1}{c|}{37} & \multicolumn{1}{c|}{0} & \multicolumn{1}{c|}{28} & 12 \\ \hline
\multicolumn{1}{|c|}{\textbf{Venus/12m/30pc}} & \multicolumn{1}{c|}{81} & \multicolumn{1}{c|}{37} & \multicolumn{1}{c|}{54} & \multicolumn{1}{c|}{83} & \multicolumn{1}{c|}{0} & \multicolumn{1}{c|}{85} & \multicolumn{1}{c|}{88} & \multicolumn{1}{c|}{0} & \multicolumn{1}{c|}{81} & 38 \\ \hline
\multicolumn{1}{|c|}{\textbf{Venus/12m/40pc}} & \multicolumn{1}{c|}{123} & \multicolumn{1}{c|}{59} & \multicolumn{1}{c|}{75} & \multicolumn{1}{c|}{125} & \multicolumn{1}{c|}{0} & \multicolumn{1}{c|}{128} & \multicolumn{1}{c|}{130} & \multicolumn{1}{c|}{0} & \multicolumn{1}{c|}{123} & 59 \\ \hline
\multicolumn{1}{|c|}{\textbf{Venus/12m/50pc}} & \multicolumn{1}{c|}{164} & \multicolumn{1}{c|}{80} & \multicolumn{1}{c|}{95} & \multicolumn{1}{c|}{166} & \multicolumn{1}{c|}{5} & \multicolumn{1}{c|}{166} & \multicolumn{1}{c|}{151} & \multicolumn{1}{c|}{21} & \multicolumn{1}{c|}{165} & 80 \\ \hline
\multicolumn{1}{|c|}{\textbf{Mars/12m/10pc}} & \multicolumn{1}{c|}{0} & \multicolumn{1}{c|}{0} & \multicolumn{1}{c|}{0} & \multicolumn{1}{c|}{0} & \multicolumn{1}{c|}{0} & \multicolumn{1}{c|}{0} & \multicolumn{1}{c|}{0} & \multicolumn{1}{c|}{0} & \multicolumn{1}{c|}{0} & 0 \\ \hline
\multicolumn{1}{|c|}{\textbf{Mars/12m/20pc}} & \multicolumn{1}{c|}{0} & \multicolumn{1}{c|}{0} & \multicolumn{1}{c|}{0} & \multicolumn{1}{c|}{0} & \multicolumn{1}{c|}{0} & \multicolumn{1}{c|}{0} & \multicolumn{1}{c|}{0} & \multicolumn{1}{c|}{0} & \multicolumn{1}{c|}{0} & 0 \\ \hline
\multicolumn{1}{|c|}{\textbf{Mars/12m/30pc}} & \multicolumn{1}{c|}{59} & \multicolumn{1}{c|}{0} & \multicolumn{1}{c|}{0} & \multicolumn{1}{c|}{0} & \multicolumn{1}{c|}{0} & \multicolumn{1}{c|}{0} & \multicolumn{1}{c|}{0} & \multicolumn{1}{c|}{0} & \multicolumn{1}{c|}{0} & 0 \\ \hline
\multicolumn{1}{|c|}{\textbf{Mars/12m/40pc}} & \multicolumn{1}{c|}{121} & \multicolumn{1}{c|}{0} & \multicolumn{1}{c|}{64} & \multicolumn{1}{c|}{1} & \multicolumn{1}{c|}{0} & \multicolumn{1}{c|}{0} & \multicolumn{1}{c|}{0} & \multicolumn{1}{c|}{0} & \multicolumn{1}{c|}{0} & 0 \\ \hline
\multicolumn{1}{|c|}{\textbf{Mars/12m/50pc}} & \multicolumn{1}{c|}{169} & \multicolumn{1}{c|}{0} & \multicolumn{1}{c|}{94} & \multicolumn{1}{c|}{23} & \multicolumn{1}{c|}{28} & \multicolumn{1}{c|}{53} & \multicolumn{1}{c|}{0} & \multicolumn{1}{c|}{68} & \multicolumn{1}{c|}{0} & 79 \\ \hline
\multicolumn{11}{|c|}{{\ul \textbf{Water (0.94 microns)}}} \\ \hline
\multicolumn{1}{|c|}{\textbf{Mercury/6m/10pc}} & \multicolumn{1}{c|}{0} & \multicolumn{1}{c|}{0} & \multicolumn{1}{c|}{0} & \multicolumn{1}{c|}{0} & \multicolumn{1}{c|}{0} & \multicolumn{1}{c|}{0} & \multicolumn{1}{c|}{0} & \multicolumn{1}{c|}{0} & \multicolumn{1}{c|}{0} & 0 \\ \hline
\multicolumn{1}{|c|}{\textbf{Mercury/6m/20pc}} & \multicolumn{1}{c|}{92} & \multicolumn{1}{c|}{92} & \multicolumn{1}{c|}{91} & \multicolumn{1}{c|}{108} & \multicolumn{1}{c|}{93} & \multicolumn{1}{c|}{91} & \multicolumn{1}{c|}{92} & \multicolumn{1}{c|}{91} & \multicolumn{1}{c|}{93} & 98 \\ \hline
\multicolumn{1}{|c|}{\textbf{Mercury/6m/30pc}} & \multicolumn{1}{c|}{212} & \multicolumn{1}{c|}{215} & \multicolumn{1}{c|}{231} & \multicolumn{1}{c|}{230} & \multicolumn{1}{c|}{230} & \multicolumn{1}{c|}{217} & \multicolumn{1}{c|}{213} & \multicolumn{1}{c|}{212} & \multicolumn{1}{c|}{219} & 230 \\ \hline
\multicolumn{1}{|c|}{\textbf{Mercury/6m/40pc}} & \multicolumn{1}{c|}{*} & \multicolumn{1}{c|}{*} & \multicolumn{1}{c|}{*} & \multicolumn{1}{c|}{*} & \multicolumn{1}{c|}{*} & \multicolumn{1}{c|}{*} & \multicolumn{1}{c|}{*} & \multicolumn{1}{c|}{*} & \multicolumn{1}{c|}{*} & * \\ \hline
\multicolumn{1}{|c|}{\textbf{Mercury/6m/50pc}} & \multicolumn{1}{c|}{*} & \multicolumn{1}{c|}{*} & \multicolumn{1}{c|}{*} & \multicolumn{1}{c|}{*} & \multicolumn{1}{c|}{*} & \multicolumn{1}{c|}{*} & \multicolumn{1}{c|}{*} & \multicolumn{1}{c|}{*} & \multicolumn{1}{c|}{*} & * \\ \hline
\multicolumn{1}{|c|}{\textbf{Venus/6m/10pc}} & \multicolumn{1}{c|}{57} & \multicolumn{1}{c|}{25} & \multicolumn{1}{c|}{42} & \multicolumn{1}{c|}{59} & \multicolumn{1}{c|}{0} & \multicolumn{1}{c|}{61} & \multicolumn{1}{c|}{64} & \multicolumn{1}{c|}{0} & \multicolumn{1}{c|}{57} & 26 \\ \hline
\multicolumn{1}{|c|}{\textbf{Venus/6m/20pc}} & \multicolumn{1}{c|}{162} & \multicolumn{1}{c|}{78} & \multicolumn{1}{c|}{94} & \multicolumn{1}{c|}{164} & \multicolumn{1}{c|}{4} & \multicolumn{1}{c|}{165} & \multicolumn{1}{c|}{150} & \multicolumn{1}{c|}{20} & \multicolumn{1}{c|}{162} & 79 \\ \hline
\multicolumn{1}{|c|}{\textbf{Venus/6m/30pc}} & \multicolumn{1}{c|}{274} & \multicolumn{1}{c|}{132} & \multicolumn{1}{c|}{147} & \multicolumn{1}{c|}{273} & \multicolumn{1}{c|}{63} & \multicolumn{1}{c|}{217} & \multicolumn{1}{c|}{202} & \multicolumn{1}{c|}{76} & \multicolumn{1}{c|}{275} & 132 \\ \hline
\multicolumn{1}{|c|}{\textbf{Venus/6m/40pc}} & \multicolumn{1}{c|}{366} & \multicolumn{1}{c|}{208} & \multicolumn{1}{c|}{219} & \multicolumn{1}{c|}{355} & \multicolumn{1}{c|}{213} & \multicolumn{1}{c|}{294} & \multicolumn{1}{c|}{276} & \multicolumn{1}{c|}{214} & \multicolumn{1}{c|}{366} & 208 \\ \hline
\multicolumn{1}{|c|}{\textbf{Venus/6m/50pc}} & \multicolumn{1}{c|}{*} & \multicolumn{1}{c|}{*} & \multicolumn{1}{c|}{*} & \multicolumn{1}{c|}{*} & \multicolumn{1}{c|}{*} & \multicolumn{1}{c|}{*} & \multicolumn{1}{c|}{*} & \multicolumn{1}{c|}{*} & \multicolumn{1}{c|}{*} & * \\ \hline
\multicolumn{1}{|c|}{\textbf{Mars/6m/10pc}} & \multicolumn{1}{c|}{0} & \multicolumn{1}{c|}{0} & \multicolumn{1}{c|}{0} & \multicolumn{1}{c|}{0} & \multicolumn{1}{c|}{0} & \multicolumn{1}{c|}{0} & \multicolumn{1}{c|}{0} & \multicolumn{1}{c|}{0} & \multicolumn{1}{c|}{0} & 0 \\ \hline
\multicolumn{1}{|c|}{\textbf{Mars/6m/20pc}} & \multicolumn{1}{c|}{167} & \multicolumn{1}{c|}{0} & \multicolumn{1}{c|}{92} & \multicolumn{1}{c|}{22} & \multicolumn{1}{c|}{27} & \multicolumn{1}{c|}{52} & \multicolumn{1}{c|}{0} & \multicolumn{1}{c|}{65} & \multicolumn{1}{c|}{0} & 77 \\ \hline
\multicolumn{1}{|c|}{\textbf{Mars/6m/30pc}} & \multicolumn{1}{c|}{238} & \multicolumn{1}{c|}{32} & \multicolumn{1}{c|}{162} & \multicolumn{1}{c|}{64} & \multicolumn{1}{c|}{82} & \multicolumn{1}{c|}{97} & \multicolumn{1}{c|}{30} & \multicolumn{1}{c|}{136} & \multicolumn{1}{c|}{0} & 179 \\ \hline
\multicolumn{1}{|c|}{\textbf{Mars/6m/40pc}} & \multicolumn{1}{c|}{289} & \multicolumn{1}{c|}{72} & \multicolumn{1}{c|}{235} & \multicolumn{1}{c|}{104} & \multicolumn{1}{c|}{147} & \multicolumn{1}{c|}{140} & \multicolumn{1}{c|}{78} & \multicolumn{1}{c|}{185} & \multicolumn{1}{c|}{33} & 252 \\ \hline
\multicolumn{1}{|c|}{\textbf{Mars/6m/50pc}} & \multicolumn{1}{c|}{337} & \multicolumn{1}{c|}{115} & \multicolumn{1}{c|}{302} & \multicolumn{1}{c|}{149} & \multicolumn{1}{c|}{242} & \multicolumn{1}{c|}{190} & \multicolumn{1}{c|}{145} & \multicolumn{1}{c|}{251} & \multicolumn{1}{c|}{82} & 340 \\ \hline
\multicolumn{1}{|c|}{\textbf{Mercury/12m/10pc}} & \multicolumn{1}{c|}{0} & \multicolumn{1}{c|}{0} & \multicolumn{1}{c|}{0} & \multicolumn{1}{c|}{0} & \multicolumn{1}{c|}{0} & \multicolumn{1}{c|}{0} & \multicolumn{1}{c|}{0} & \multicolumn{1}{c|}{0} & \multicolumn{1}{c|}{0} & 0 \\ \hline
\multicolumn{1}{|c|}{\textbf{Mercury/12m/20pc}} & \multicolumn{1}{c|}{0} & \multicolumn{1}{c|}{0} & \multicolumn{1}{c|}{0} & \multicolumn{1}{c|}{0} & \multicolumn{1}{c|}{0} & \multicolumn{1}{c|}{0} & \multicolumn{1}{c|}{0} & \multicolumn{1}{c|}{0} & \multicolumn{1}{c|}{0} & 0 \\ \hline
\multicolumn{1}{|c|}{\textbf{Mercury/12m/30pc}} & \multicolumn{1}{c|}{15} & \multicolumn{1}{c|}{18} & \multicolumn{1}{c|}{17} & \multicolumn{1}{c|}{15} & \multicolumn{1}{c|}{10} & \multicolumn{1}{c|}{5} & \multicolumn{1}{c|}{13} & \multicolumn{1}{c|}{16} & \multicolumn{1}{c|}{18} & 17 \\ \hline
\multicolumn{1}{|c|}{\textbf{Mercury/12m/40pc}} & \multicolumn{1}{c|}{92} & \multicolumn{1}{c|}{92} & \multicolumn{1}{c|}{91} & \multicolumn{1}{c|}{108} & \multicolumn{1}{c|}{93} & \multicolumn{1}{c|}{91} & \multicolumn{1}{c|}{92} & \multicolumn{1}{c|}{91} & \multicolumn{1}{c|}{93} & 98 \\ \hline
\multicolumn{1}{|c|}{\textbf{Mercury/12m/50pc}} & \multicolumn{1}{c|}{150} & \multicolumn{1}{c|}{150} & \multicolumn{1}{c|}{159} & \multicolumn{1}{c|}{168} & \multicolumn{1}{c|}{163} & \multicolumn{1}{c|}{150} & \multicolumn{1}{c|}{150} & \multicolumn{1}{c|}{150} & \multicolumn{1}{c|}{150} & 164 \\ \hline
\multicolumn{1}{|c|}{\textbf{Venus/12m/10pc}} & \multicolumn{1}{c|}{0} & \multicolumn{1}{c|}{0} & \multicolumn{1}{c|}{0} & \multicolumn{1}{c|}{0} & \multicolumn{1}{c|}{0} & \multicolumn{1}{c|}{0} & \multicolumn{1}{c|}{0} & \multicolumn{1}{c|}{0} & \multicolumn{1}{c|}{0} & 0 \\ \hline
\multicolumn{1}{|c|}{\textbf{Venus/12m/20pc}} & \multicolumn{1}{c|}{57} & \multicolumn{1}{c|}{25} & \multicolumn{1}{c|}{42} & \multicolumn{1}{c|}{59} & \multicolumn{1}{c|}{0} & \multicolumn{1}{c|}{61} & \multicolumn{1}{c|}{64} & \multicolumn{1}{c|}{0} & \multicolumn{1}{c|}{57} & 26 \\ \hline
\multicolumn{1}{|c|}{\textbf{Venus/12m/30pc}} & \multicolumn{1}{c|}{111} & \multicolumn{1}{c|}{53} & \multicolumn{1}{c|}{69} & \multicolumn{1}{c|}{113} & \multicolumn{1}{c|}{0} & \multicolumn{1}{c|}{116} & \multicolumn{1}{c|}{118} & \multicolumn{1}{c|}{0} & \multicolumn{1}{c|}{111} & 53 \\ \hline
\multicolumn{1}{|c|}{\textbf{Venus/12m/40pc}} & \multicolumn{1}{c|}{162} & \multicolumn{1}{c|}{78} & \multicolumn{1}{c|}{94} & \multicolumn{1}{c|}{164} & \multicolumn{1}{c|}{4} & \multicolumn{1}{c|}{165} & \multicolumn{1}{c|}{150} & \multicolumn{1}{c|}{20} & \multicolumn{1}{c|}{162} & 79 \\ \hline
\multicolumn{1}{|c|}{\textbf{Venus/12m/50pc}} & \multicolumn{1}{c|}{215} & \multicolumn{1}{c|}{105} & \multicolumn{1}{c|}{120} & \multicolumn{1}{c|}{216} & \multicolumn{1}{c|}{31} & \multicolumn{1}{c|}{190} & \multicolumn{1}{c|}{175} & \multicolumn{1}{c|}{47} & \multicolumn{1}{c|}{216} & 105 \\ \hline
\multicolumn{1}{|c|}{\textbf{Mars/12m/10pc}} & \multicolumn{1}{c|}{0} & \multicolumn{1}{c|}{0} & \multicolumn{1}{c|}{0} & \multicolumn{1}{c|}{0} & \multicolumn{1}{c|}{0} & \multicolumn{1}{c|}{0} & \multicolumn{1}{c|}{0} & \multicolumn{1}{c|}{0} & \multicolumn{1}{c|}{0} & 0 \\ \hline
\multicolumn{1}{|c|}{\textbf{Mars/12m/20pc}} & \multicolumn{1}{c|}{0} & \multicolumn{1}{c|}{0} & \multicolumn{1}{c|}{0} & \multicolumn{1}{c|}{0} & \multicolumn{1}{c|}{0} & \multicolumn{1}{c|}{0} & \multicolumn{1}{c|}{0} & \multicolumn{1}{c|}{0} & \multicolumn{1}{c|}{0} & 0 \\ \hline
\multicolumn{1}{|c|}{\textbf{Mars/12m/30pc}} & \multicolumn{1}{c|}{105} & \multicolumn{1}{c|}{0} & \multicolumn{1}{c|}{45} & \multicolumn{1}{c|}{0} & \multicolumn{1}{c|}{0} & \multicolumn{1}{c|}{0} & \multicolumn{1}{c|}{0} & \multicolumn{1}{c|}{0} & \multicolumn{1}{c|}{0} & 0 \\ \hline
\multicolumn{1}{|c|}{\textbf{Mars/12m/40pc}} & \multicolumn{1}{c|}{167} & \multicolumn{1}{c|}{0} & \multicolumn{1}{c|}{92} & \multicolumn{1}{c|}{22} & \multicolumn{1}{c|}{27} & \multicolumn{1}{c|}{52} & \multicolumn{1}{c|}{0} & \multicolumn{1}{c|}{65} & \multicolumn{1}{c|}{0} & 77 \\ \hline
\multicolumn{1}{|c|}{\textbf{Mars/12m/50pc}} & \multicolumn{1}{c|}{210} & \multicolumn{1}{c|}{11} & \multicolumn{1}{c|}{126} & \multicolumn{1}{c|}{44} & \multicolumn{1}{c|}{55} & \multicolumn{1}{c|}{76} & \multicolumn{1}{c|}{7} & \multicolumn{1}{c|}{112} & \multicolumn{1}{c|}{0} & 130 \\ \hline
\end{longtable}

\bibliography{sample631}{}
\bibliographystyle{aasjournal}

\end{document}